\documentclass[acmsmall,screen]{acmart}

\AtBeginDocument{%
  \providecommand\BibTeX{{%
    Bib\TeX}}}

\acmPrice{}
\acmISBN{978-1-4503-XXXX-X/18/06}

\usepackage{multirow}
\usepackage[ruled,linesnumbered]{algorithm2e}
\usepackage{amsmath}
\usepackage{subcaption}
\usepackage{color,xcolor,colortbl}
\usepackage{enumitem}
\usepackage{graphicx}
\usepackage{tcolorbox}
\usepackage{bbding}
\usepackage{microtype}
\usepackage{tikz}   % For drawing
\usepackage{setspace}
\usepackage{bm}
\begin{document}

% \title{Automated Generation of Bug Reproduction Scripts from Issue Descriptions Using Code Agents}
\title{AEGIS: An Agent-based Framework for General Bug Reproduction from Issue Descriptions}

\author{Xinchen Wang}
\authornote{Work done during an internship at ByteDance.}
\email{200111115@stu.hit.edu.cn}
\affiliation{
    \institution{Haribin Institute of Technology, Shenzhen}
    \city{Shenzhen}
    \country{China}
}

\author{Pengfei Gao}
\email{gaopengfei.se@bytedance.com}
\affiliation{
    \institution{ByteDance}
    \city{Beijing}
    \country{China}
}

\author{Xiangxin Meng}
\email{mengxiangxin.1219@bytedance.com}
\affiliation{
    \institution{ByteDance}
    \city{Beijing}
    \country{China}
}

\author{Chao Peng}
\authornote{Corresponding authors.}
\email{pengchao.x@bytedance.com}
\affiliation{
    \institution{ByteDance}
    \city{Beijing}
    \country{China}
}

\author{Ruida Hu}
\authornotemark[1]
\email{200111107@stu.hit.edu.cn}
\affiliation{
    \institution{Haribin Institute of Technology, Shenzhen}
    \city{Shenzhen}
    \country{China}
}

\author{Yun Lin}
\email{lin_yun@sjtu.edu.cn}
\affiliation{
    \institution{Shanghai Jiao Tong University}
    \city{Shanghai}
    \country{China}
}

\author{Cuiyun Gao}
\authornotemark[2]
\email{gaocuiyun@hit.edu.cn}
\affiliation{
    \institution{Haribin Institute of Technology, Shenzhen}
    \city{Shenzhen}
    \country{China}
}

%\author{Anonymous Author(s)}
\renewcommand{\shortauthors}{Anonymous}
\definecolor{darkgreen}{rgb}{0,0.5,0}
\newcommand{\framework}{AEGIS\xspace}
\newcommand{\moduleA}{concise context construction module\xspace}
\newcommand{\moduleB}{FSM-based multi-feedback optimization module\xspace}
\newcommand{\ie}{\textit{i}.\textit{e}.\xspace}
\newcommand{\metricA}{$\frac{F \rightarrow P}{F \rightarrow X}$\xspace}
\newcommand{\metricB}{$\frac{F \rightarrow P}{X \rightarrow P}$\xspace}
\newcommand{\greendownarrow}{\textcolor{darkgreen}{$\downarrow$}}
\newcommand{\reduparrow}{\textcolor{red}{$\uparrow$}}
\newcommand{\wxc}[1]{\textcolor{brown}{{#1}}}
\newcommand{\cdw}[1]{\textcolor{purple}{{#1}}}
\newcommand{\wen}[1]{\textcolor{blue}{{#1}}}

\begin{abstract}
\label{sec:abstract}
 In software maintenance, bug reproduction is essential for effective fault localization and repair. Manually writing reproduction scripts is a time-consuming task with high requirements for developers. Hence, automation of bug reproduction has increasingly attracted attention from researchers and practitioners. However, the existing studies on bug reproduction are generally limited to specific bug types such as program crashes, and hard to be applied to general bug reproduction. In this paper, considering the superior performance of agent-based methods in code intelligence tasks, we focus on designing an agent-based framework for the task. 
 
 Directly employing agents would lead to limited bug reproduction performance, due to entangled subtasks, lengthy retrieved context, and unregulated actions. To mitigate the challenges, we propose an \textbf{A}utomated g\textbf{E}neral bu\textbf{G} reproduct\textbf{I}on \textbf{S}cripts generation framework, named \textbf{\framework}, which is the first agent-based framework for the task. \framework mainly contains two modules: (1) A \moduleA, which aims to guide the code agent in extracting structured information from issue descriptions, identifying issue-related code with detailed explanations, and integrating these elements to construct the concise context; (2) A \moduleB to further regulate the behavior of the code agent within the finite state machine (FSM), ensuring a controlled and efficient script generation process based on multi-dimensional feedback. Extensive experiments on the public benchmark dataset show that \framework outperforms the state-of-the-art baseline by 23.0\% in $F \rightarrow P$ metric. In addition, the bug reproduction scripts generated by \framework can improve the relative resolved rate of Agentless by 12.5\%. 


\end{abstract}




\maketitle

\section{INTRODUCTION}
\label{sec:introduction}
Software testing~\cite{softwaretest1, softwaretest2, softwaretest3} plays a crucial part in the software development lifecycle, which involves executing tests to identify and verify functionalities bugs. Existing automated test generation techniques~\cite{auto1, auto2, auto3, DBLP:journals/pacmse/Yuan0DW00L24, DBLP:conf/sigsoft/ChenHZHDY24, DBLP:journals/tse/SchaferNET24} are capable of producing high-quality unit tests for specified functionalities, thereby reducing the repairing effort and enhancing development efficiency. 
 
 Reproducing general bugs from issue descriptions is very essential. Numerous tests in project repositories originate from issue reports \cite{DBLP:conf/icse/KangYY23}, aiming to prevent their recurrence, which has become a standard practice in workflows. 
 Multiple studies have concluded that generating bug reproduction scripts from issues closely aligns with developers' requirements~\cite{need1,need2} and enhances the performance of automated debugging tools~\cite{autoreproduce1,autoreproduce2}. The dynamic execution information from reproduction scripts~\cite{dynamic} aids in more precise bug localization and targeted solutions. However, manually constructing these scripts is time-consuming and demands strong multilingual understanding and programming skills. Meanwhile, existing reproduction techniques~\cite{autoreproduce1,autoreproduce2} primarily focus on addressing crashes and are not applicable to general bugs described in issue descriptions. As a result, the task of reproducing general bugs from issue descriptions has gained increasing attention.
 
 Program repair-based code agents demonstrate the best performance in general bug reproduction; however, optimized agent-based methods for this task are lacking. Considering the powerful context understanding of large language models (LLMs), Libro~\cite{DBLP:conf/icse/KangYY23} leverages LLMs and prompt engineering~\cite{promptengineering} to achieve effective outcomes in the bug reproduction task. However, Libro cannot dynamically execute and modify reproduction scripts, limiting its performance. Code agents~\cite{DBLP:journals/corr/abs-2406-01304, liu2024marscodeagentainativeautomated, DBLP:journals/corr/abs-2403-17134}, as intelligent systems that integrate code retrieval, generation, and editing, can invoke various external tools and APIs, process execution feedback, and take subsequent actions~\cite{DBLP:journals/corr/abs-2308-10848,DBLP:journals/corr/abs-2307-07924, DBLP:conf/acl/Qiao0FLZJLC24}. Currently, code agents have demonstrated great promise in program repair~\cite{DBLP:journals/corr/abs-2405-15793, DBLP:journals/corr/abs-2404-05427}. Niels et al.~\cite{DBLP:journals/corr/abs-2406-12952} make simple prompt adjustments to existing program repair-based code agents, achieving state-of-the-art results in the general bug reproduction task. However, they do not design modules or optimize methods specifically for this task.

% \wxc{\textbf{The performance of generating problem reproduction scripts on the SWE-bench Lite~\cite{swebench} benchmark falls far below expectations.} Niels et al. evaluate the performance of multiple methods based on a subset of SWE-bench Lite data, with the best `PASS@5' method achieving only 11.5\%. However, Agentless~\cite{DBLP:journals/corr/abs-2407-01489} analyzes each instance and concludes that 54.7\% contain complete reproduction examples. Therefore, there is a great gap between the actual result of 11.5\% and the expected result of 63.4\%. To verify whether the situation is truly as Agentless claims, a reanalysis is urgently needed.}

Code agents face numerous challenges in generating reproduction scripts.
\textbf{(1) Entangled Subtasks:} During the process of bug reproduction, code agents perform multiple subtasks, including context retrieval, reproduction script execution, and multi-file editing. The frequent switching between these subtasks reduces the efficiency.
\textbf{(2) Lengthy Retrieved Context:} While retrieving context relevant to the issue description, code agents also collect a large amount of irrelevant intermediate results. An overly large context window increases the burden on the code agents~\cite{challenge2}.
\textbf{(3) Unregulated Actions:} Code agents obtain various feedbacks by invoking external tools. However, they struggle to effectively utilize these feedbacks and even take incorrect actions~\cite{challenge3}.

% (1) How can we facilitate a better understanding of problem descriptions for Code Agent? Problem descriptions on platforms like GitHub and Stack Overflow are predominantly unstructured, encompassing a wide array of information. This complexity poses a challenge to the code agent's comprehension of the problem and its ability to localize errors.
 % (2) How can we direct Code Agent to generate reproduction scripts more efficiently? Code Agent exhibits a high degree of autonomy, capable of self-optimization and adjustment. However, this autonomy also introduces potential unpredictability. The challenge lies in standardizing the code agent's behavior while preserving its flexibility.

To address the above limitations and challenges, in this paper, we propose an \textbf{A}utomated g\textbf{E}neral bu\textbf{G} reproduct\textbf{I}on \textbf{S}cripts generation framework, named \textbf{\framework}. \framework mainly contains two modules: (1) A \moduleA, which aims to extract structured information from issue descriptions, identify issue-related code with detailed explanations, and integrate these elements to construct the comprehensive context; (2) A \moduleB to further regulate the behavior of the agent within the finite state machine (FSM)~\cite{fsm}, ensuring a controlled and efficient script generation process based on multi-dimensional feedback. Extensive experiments on the public benchmark dataset show that \framework outperforms the state-of-the-art baseline by 23.0\% in $F \rightarrow P$ metric. In addition, the bug reproduction scripts generated by \framework can improve the relative resolved rate of Agentless by 12.5\%. 

In summary, the major contributions of this paper are summarized as follows:
\begin{enumerate}
 \item We are the first to focus on optimizing agent-based methods for generating general bug reproduction scripts from issue descriptions.
  % \item \wxc{We conclude that there is still great room for developing intelligent techniques for bug reproduction tasks based on a detailed reanalysis of a subset of SWE-bench Lite.} 
 \item We propose \framework, a novel agent-based general bug reproduction framework, which involves a \moduleA for better issue understanding and bug comprehension and a \moduleB for the controlled and efficient script generation process.
 \item We evaluate \framework in the popular benchmark dataset, and the results demonstrate the effectiveness of \framework in general bug reproduction. Furthermore, the bug reproduction scripts generated by \framework can enhance the performance of existing program repair tools.
\end{enumerate}

The remaining sections of this paper are organized as follows. 
% Section~\ref{sec:motivation} conducts a study on the performance of code agents in reproducing general bugs and reveals the associated challenges. 
Section~\ref{sec:motivation} reveals the challenges faced by code agents in reproducing general bugs from issue descriptions. 
Section~\ref{sec:architecture} presents the architecture of \framework, which includes two modules: a \moduleA and a \moduleB. Section~\ref{sec:evaluation} describes the experimental setup, including datasets, baselines, and experimental settings. 
Section~\ref{sec:experimental_result} presents the experimental results and analysis.
Section~\ref{sec:discussion} discusses \framework's performance against current challenges, the impact of the retrieved context on \framework, and the threats to validity. Section~\ref{sec:conclusion} concludes the paper.

\section{MOTIVATION}
\label{sec:motivation}
\begin{figure}[t]
	\centering
	\includegraphics[width=1\textwidth]{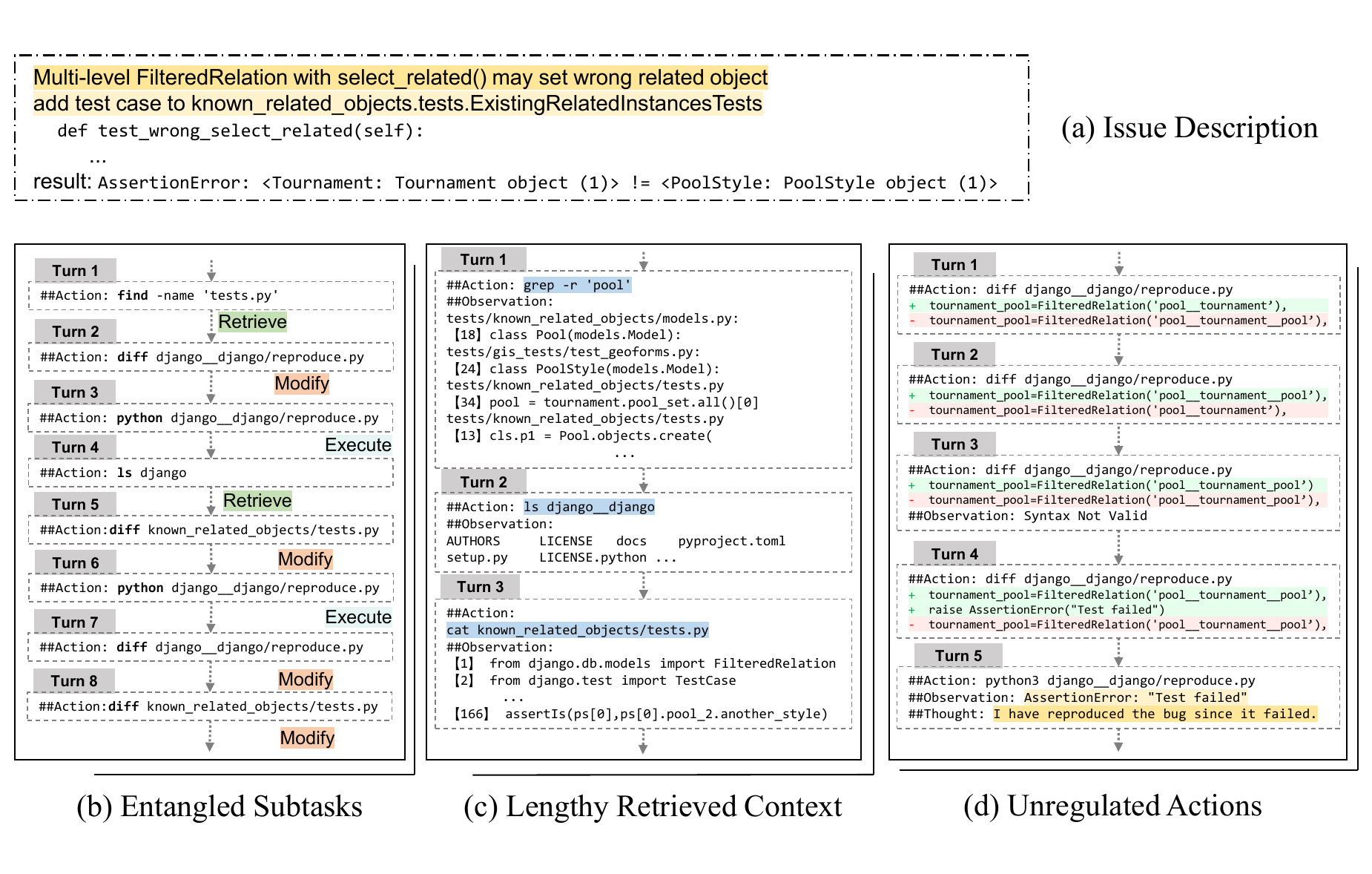}
    \caption{Examples for illustrating the challenges of code agents in general bug reproduction. Turn X represents the relative order of different turns.}
\label{fig:motivation}
\end{figure}

% In this section, we conduct a case study and analysis on the performance of code agents in reproducing issues and aim to uncover code agents' challenges and limitations in constructing bug reproduction scripts.  Since the current agent-based methods for issue reproduction are adapted from those used for program repair, both conclusions are similar.

% The general steps of code agent-based program repair methods~\cite{DBLP:journals/corr/abs-2406-01304,  DBLP:journals/corr/abs-2403-17134} are as follows: First, code agents retrieve contextual information related to the issue description by calling APIs. Based on the collected context, code agents edit and run bug reproduction scripts to obtain execution messages. By analyzing the execution messages, code agents localize bugs and generate repair patches. Finally, the code agents apply the repair patches and rerun the reproduction scripts to confirm the effectiveness of the repairs.

% MarsCode Agent, recognized as the advanced program repair method in the industry, generally follows the aforementioned steps. Therefore, we select MarsCode Agent for our case study. 

In this section, we explore and summarize the challenges faced by the code agent in the general bug reproduction task.

Since existing code agents~\cite{DBLP:journals/corr/abs-2405-15793, DBLP:journals/corr/abs-2406-01304,  DBLP:journals/corr/abs-2403-17134} are designed for program repair tasks, we design a simple AgentBaseline for bug reproduction tasks based on the existing code agent framework. AgentBaseline possesses the capability to search and view code, as well as to edit and execute bug reproduction scripts. It simultaneously performs context retrieval, script execution, and file editing, submitting the refined reproduction script. Figure~\ref{fig:motivation} illustrates the issue description and part of the reproduction process for instance \textit{`django-16408'}. Figure~\ref{fig:motivation}(a) indicates that the issue arises due to `Multi-level FilteredRelation with $select\_related()$ may set wrong related object', suggesting adding a test case in the $ExistingRelatedInstancesTests$ class and providing the current output.

\textbf{(1) Entangled Subtasks: Decouple the subtasks of context retrieval, script execution, and multi-file editing.} As shown in Figure~\ref{fig:motivation}(b), the code agent alternates between retrieving context (Turn 1, 4), editing files (Turn 2, 5), and executing the scripts (Turn 3, 6). The code agent rushes into reproduction script editing without adequately gathering context, resulting in unsuccessful attempts and necessitating further information retrieval. Additionally, the code agent has to modify \textit{tests.py} to add the test case described in the issue and create \textit{reproduce.py} as the test entry point. However, if the reproduction fails, both files need to be modified simultaneously (Turn 2, 5, 7, 8). The entanglement of these subtasks results in high costs with low efficiency.

\textbf{(2) Lengthy Retrieved Context: Extract and integrate issue-related context.} As shown in Figure~\ref{fig:motivation}(c), the code agent invokes various APIs and system commands for context retrieval. In Turn 1, `$grep -r$' is called to search for the identifier `pool' mentioned in the issue description, returning multiple results. In Turn 2, the `$ls$' command is used to list all files under the project directory. In Turn 3, `$cat$' is invoked to browse through `tests.py', which spans 166 lines. Obviously, not all retrieved context is pertinent to the issue. For instance, the output of commands like `$ls$' and `$cat$' provides minimal assistance in understanding the issue. The issue-related information is sparse and spread across a large context window, making comprehension difficult.

\textbf{(3) Unregulated Actions: Effectively utilize feedback and adopt more regulated actions.} On one hand, the code agent can leverage various external tools to obtain feedback. As shown in Figure~\ref{fig:motivation}(d), in Turn 3, after applying the patch, the built-in syntax checking tool detects and reports syntax errors. In Turn 5, the code agent executes the reproduction script and receives error messages. On the other hand, the code agent exhibits high flexibility, with its thinking and actions not being constrained. In Turn 4, after multiple unsuccessful attempts to reproduce the bug, the code agent resorts to utilizing a `raise AssertionError' statement to explicitly trigger an error. In Turn 5, the code agent mistakenly interprets the error message "Test Failed" as a successful reproduction of the bug, failing to consider the current output described in the issue and lacking external constraints on its judgment.

Based on the findings and analysis above, we propose an \textbf{A}utomated bug r\textbf{E}production scripts \textbf{G}eneration framework from \textbf{I}ssue de\textbf{S}criptions, named \textbf{\framework} in this paper. \framework decouples the subtasks of context retrieval, script execution, and multi-file editing. \framework extracts essential information from the code context and issue description, reducing the code agent's context window. Additionally, \framework takes actions based on feedback within a finite state machine framework, ensuring a controlled and effective process.

% \section{Empirical Study}
% \label{sec:empirical}
% \input{Sections/2_Empirical}

\section{PROPOSED FRAMEWORK}
\label{sec:architecture}
\begin{figure}[t]
	\centering
	\includegraphics[width=1\textwidth]{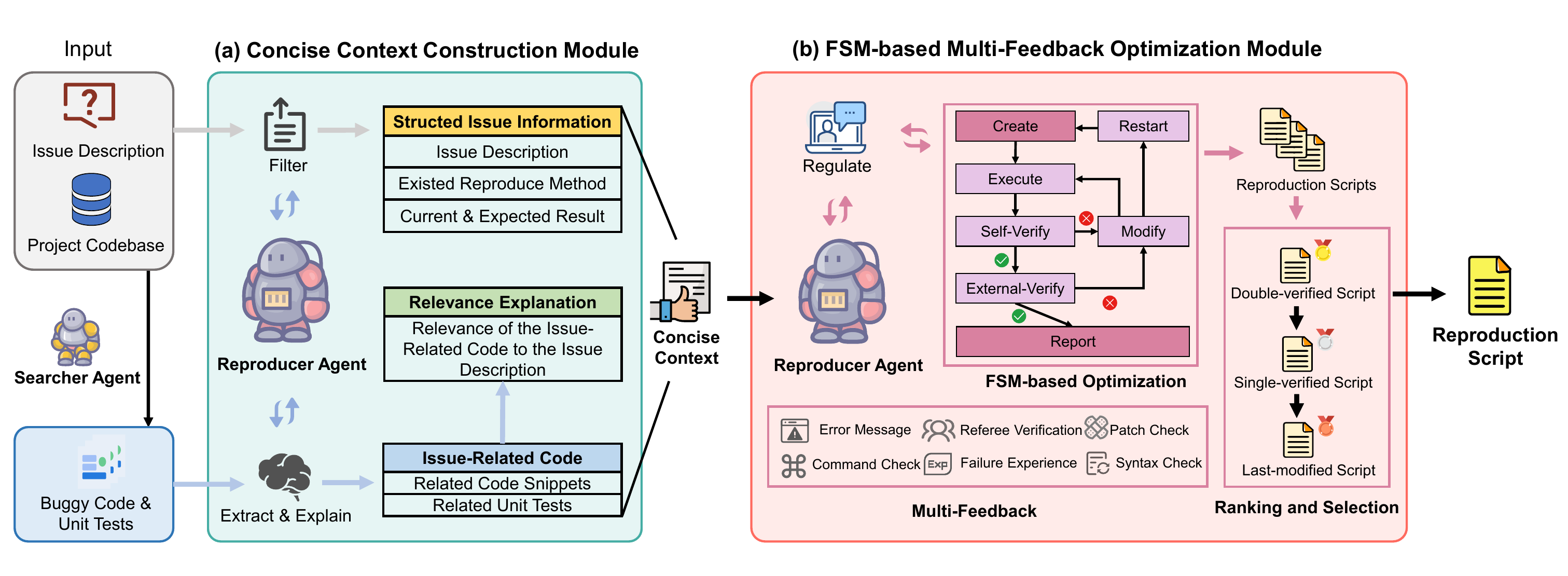}
    \caption{The architecture of \framework. \framework mainly consists of two modules driven by the Reproducer Agent: a \moduleA and a \moduleB. The buggy code and unit tests retrieved by the Searcher Agent, issue description, and the project codebase are the input of \framework. The output is the reproduction script.
    }
\label{fig:architecture}
\end{figure}

In this section, we introduce the overall framework of \framework. As shown in Figure~\ref{fig:architecture}, \framework consists of two main modules:
% driven by the Reproducer Agent: 
(1) a \moduleA for integrating essential issue-related information into a structured context, (2) a \moduleB for iterative refinement of reproduction scripts within the finite state machine process. 
The input to the \framework includes the issue description, project codebase, and the buggy code along with unit tests retrieved by the Searcher Agent.

% The Searcher Agent, based on the issue description, searches the entire repository for relevant buggy code and unit tests, providing essential context for the Reproducer Agent. Subsequently, the Reproducer Agent sequentially executes the following two modules and submits the final reproduction script:
% (1) The \moduleA aims to guide the Reproducer Agent in extracting structured information from issue descriptions, identifying issue-related code with detailed explanations, and integrating these elements to construct the core context.
% (2) The \moduleB aims to regulate the behavior of the Reproducer Agent within the finite state machine(FSM), ensuring a controlled and efficient script generation process. Multi-dimensional feedback information serves as transition conditions, allowing the Reproducer Agent to take appropriate actions in various states and submit the final reproduction script based on priority rules.

\subsection{Concise Context Construction Module}

The \moduleA aims to construct structured information from unstructured issue descriptions and extract issue-related code snippets from extensive contexts.
% , integrating these elements into the concise context. 

\subsubsection{Pilot Experience}
% \subsubsection{Structured Issue Information}

Based on the following observations and considerations, we find that directly integrating the issue descriptions and the context retrieved by the Searcher Agent as input is not appropriate: (1) Different users pose issues with distinct personal styles and inconsistent formatting. Additionally, issue descriptions contain varied information, including current results, expected results, reproduction methods, runtime environments, tool versions, and user comments. Extraneous and unstructured content may hinder the Reproducer Agent's comprehension of the issue and distract it from bug reproduction. (2) The Searcher Agent retrieves extensive code contexts based on the issue description. However, many contexts are intermediate results, such as outputs from commands `\textit{ls}' and `\textit{find}' as shown in Figure~\ref{fig:motivation}(c). These contexts may not be relevant to the issue but can greatly expand the Reproducer Agent's context window. Additionally, the absence of explanations linking the code contexts to the issue may confuse the Reproducer Agent in utilizing these contexts for bug reproduction.

% \subsubsection{Issue-Related Code and Relevance Explanation}
\subsubsection{Concise Context Construction}

\begin{figure}[t]
	\centering
	\includegraphics[width=\textwidth]{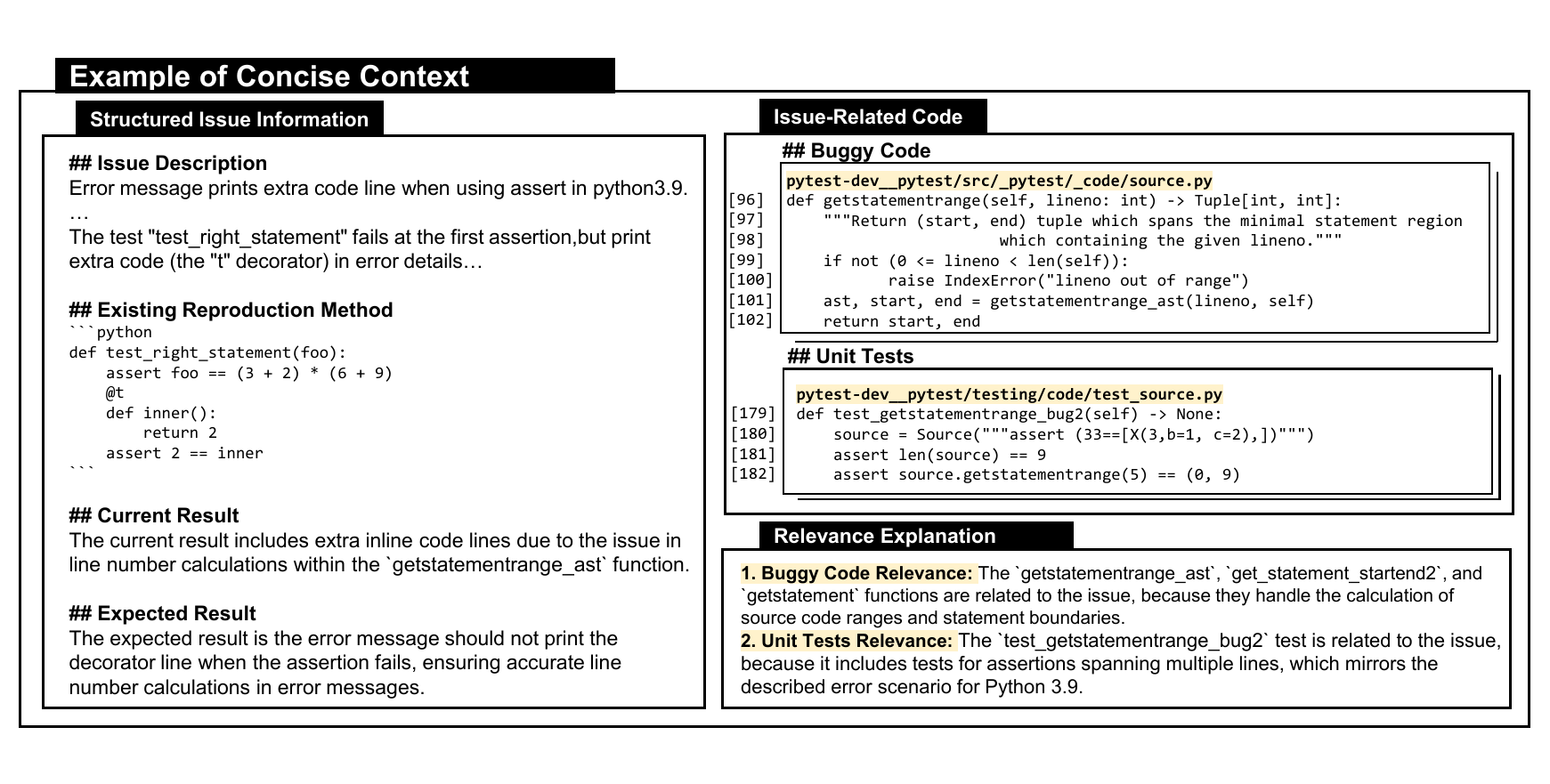}
    \caption{The example of the concise context, including the structured issue information, issue-related code, and the relevance explanation.}
\label{fig:example}
\end{figure}

Based on the above 
% considerations
findings, we 
% instruct the Reproducer Agent to 
extract three types of information:
% from the issue description: 
structured issue information, issue-related code, and relevance explanation.
Figure~\ref{fig:example} illustrates an example of the concise context.
% existing reproduction methods, current results, and expected results. 
% Additionally, we 
% % instruct the Reproducer Agent to 
% % distill 
% \wxc{extract}
% and explain the contexts, ultimately providing two types of information: issue-related code (buggy code and unit tests closely linked to the issue description) and relevance explanations (specific connections between the issue-related code and the issue description).

% Figure~\ref{fig:example} illustrates an example of the concise context, which encompasses three levels of information: 
(1) \textbf{Structured Issue Information}: comprising the issue description, existing reproduction methods, current result, and expected result.
Existing reproduction methods can reduce the effort required to generate the reproduction script, while current and expected results assist the Reproducer Agent in validation and optimization. As shown in Figure~\ref{fig:example}, the existing reproduction method includes a comprehensive unit test example, `$test\_right\_statement$`. The current result indicates that an additional line of code is being output, whereas the expected result specifies that this line should not be printed. 

(2) \textbf{Issue-Related Code}: buggy code and unit tests associated with the issue description. As illustrated in Figure~\ref{fig:example}, the buggy code and unit tests include the issue-related classes or methods, as well as the corresponding files.

(3) \textbf{Relevance Explanation}: a detailed explanation of the relevance between the issue-related code and the issue. As shown in Figure~\ref{fig:example}, this information explains the relevance between the `$getstatementrange\_ast$' and  `$test\_getstatementrange\_bug2$' and the bug. 

This concise context alleviates the cognitive burden on the Reproducer Agent, deepens its grasp of the issue description, and enables more precise reproduction.

\subsection{FSM-based Multi-Feedback Optimization Module}
The \moduleB
% , utilizing the finite state machine framework, 
leverages multi-dimensional feedback to regulate the Reproducer Agent through iterative optimization. This process results in the generation of a ranked reproduction script.
% ultimately submitting a ranked reproduction script.

\subsubsection{Initialization of Finite State Machine}

A finite state machine (FSM)~\cite{fsm} is a computational model used to represent a system with a finite number of states and the transitions between these states. FSMs are fundamental in various fields, including compiler design, network protocols, and automation systems, where they are used to model and manage state-based behavior efficiently.

Code agents struggle with complex constraints and often deviate from planned execution in bug reproduction tasks as illustrated in Figure~\ref{fig:motivation}(d), leading to ineffective scripts and reduced performance. In the FSM process, code agents make decisions based on the current state and execute predefined state transitions according to feedback, ensuring a more controlled reproduction process.

In this paper, the \moduleB is a 5-tuple $(Q, \Sigma, \delta, q_{0}, F) $, comprising the following components:
\begin{itemize}
    \item \textbf{State Set $Q$}: A finite set of states defines the transitions within the FSM. For example, the state set is $Q = \{q_{0}, q_{1}, \dots, q_{n}\}$. Each state $q_{i} \in Q$ possesses distinct attributes and is capable of executing different actions.
    \item \textbf{Input Alphabet $\Sigma$}: A finite set of symbols represents all possible inputs to the FSM. For instance, the input alphabet can be $\Sigma = \{a, b, c\}$. Input $a \in \Sigma$ can include not only characters but also events, such as input signals or timer expirations, as well as conditions like `$temperature > 100$' or `$if (flag == true)$'.
    \item \textbf{Transition function $\delta$}: 
    This defines the transition of the FSM
    % Defines how the FSM transits 
    from one state to another based on an input $a \in \Sigma$, which is typically represented as a mapping $\delta: Q \times \Sigma \rightarrow Q$. 
    \item \textbf{Initial State $q_{0} \in Q$}: The state from which the FSM starts.
    \item \textbf{Accepting State Set $F \subseteq Q$}: It is a set of states where the FSM accepts the input. 
\end{itemize}

% FSMs offer simplicity in design, clarity in state management, and efficient handling of various states and transitions, making them ideal for structured process control.

\subsubsection{FSM-based Multi-Feedback Optimization}

\begin{figure}[t]
	\centering
	\includegraphics[width=\textwidth]{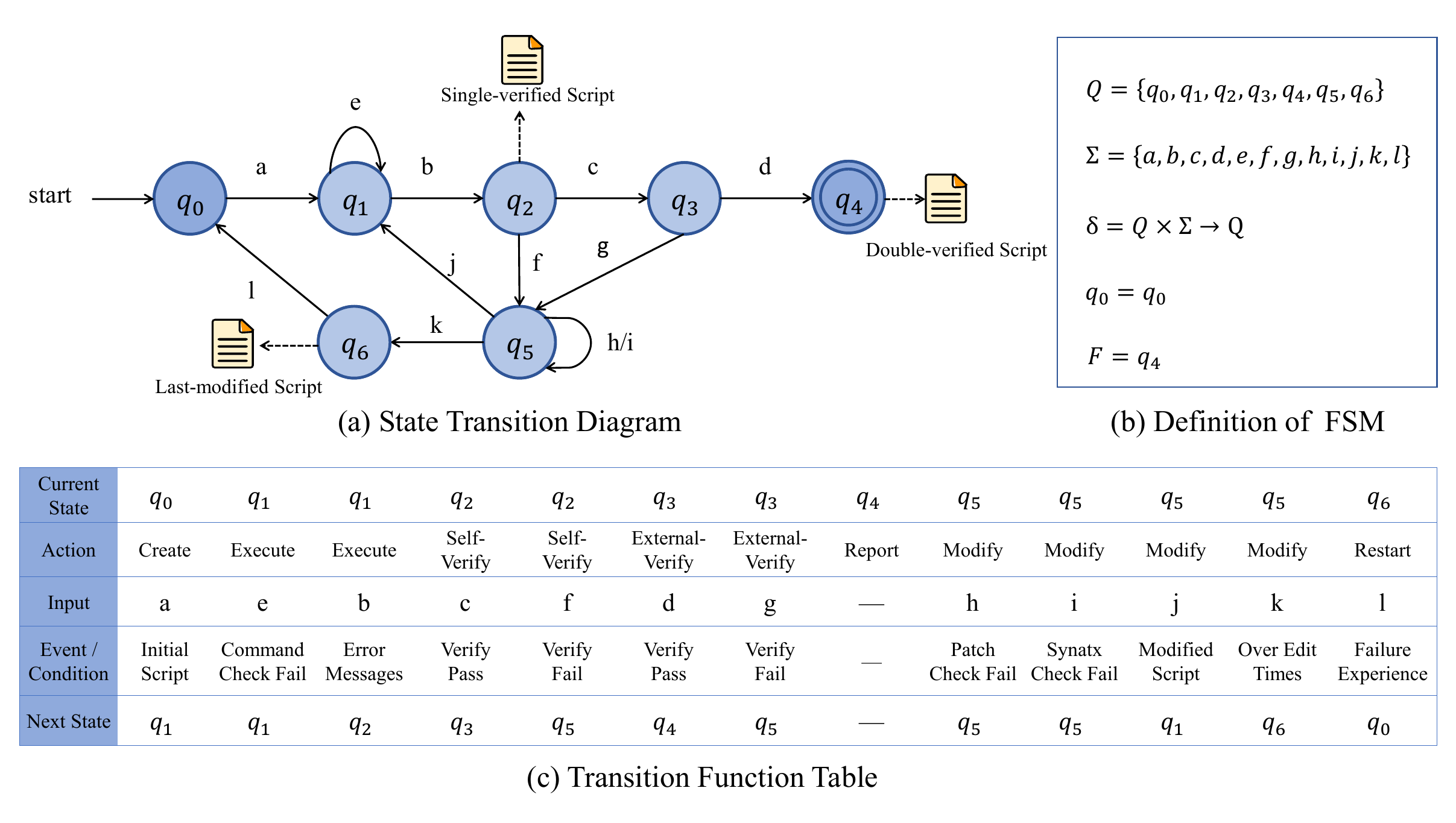}
    \caption{The architecture of our designed finite state machine (FSM). (a), (b), and (c) illustrate the state transition diagram, definition of FSM, and transition function table of the designed FSM, respectively. }
\label{fig:fsm}
\end{figure}

As illustrated in Figure~\ref{fig:fsm}, we propose a FSM-based process to
% utilize the process of FSM to 
direct the Reproducer Agent in generating bug reproduction scripts. 

\begin{itemize}
    \item \textbf{State Set $Q = \{q_0, q_1, q_2, q_3, q_4, q_5, q_6\}$} denotes the various states of the Reproducer Agent, including \textit{Create, Execute, Self-Verify, External-Verify, Report, Modify} and \textit{Restart}, respectively.
    \item \textbf{Input Alphabet $\Sigma = \{a, b, c, d, e, f, g, h, i, j, k, l\}$} represents the different events and conditions for the Reproducer Agent, including \textit{Command Check, Error Message, Syntax Check, Referee Verification, Patch Check, Failure Experience}, and others, as illustrated in Figure~\ref{fig:architecture}.
    \item \textbf{Initial State $q_0$}\textit{(Create)} indicates that the Reproducer Agent starts by creating reproduction scripts based on the constructed concise context.
    \item \textbf{Accepting State $q_4$}\textit{(Report)} indicates that the Reproducer Agent has finalized the reproduction scripts and reported success.
    \item \textbf{Transition function $\delta$} describes how the Reproducer Agent transitions between states in response to various events and conditions.
    \begin{description}[leftmargin=.5in]
        \item[$\bm{\delta(q_0, a) = q_1}$:] In \textit{(Create)} state $q_0$, the Reproducer Agent generates the \textbf{initial reproduction script}($a$) based on the constructed concise context. Then, the FSM transitions to state $q_1$.
        \item[$\bm{\delta(q_1, b) = q_2}$:] In \textit{(Execute)} state $q_1$, the Reproducer Agent executes the reproduction script and obtains \textbf{error messages}($b$). The dynamic execution information is essential for analyzing bug reproduction. Then, the FSM transitions to state $q_2$. 
        \item[$\bm{\delta(q_1, e) = q_1}$:] In \textit{(Execute)} state $q_1$, the Reproducer Agent is restricted to using the script execution command, called command check. File editing and context retrieval commands are not permitted to ensure process simplicity. If the \textbf{command check fails}($e$), a system prompt will be added to guide proper command usage, and the state remains $q_1$. 
        \item[$\bm{\delta(q_2, c) = q_3}$:] In \textit{(Self-Verify)} state $q_2$, the Reproducer Agent analyzes the error messages based on the issue description and determines whether the error message reproduces the issue. If \textbf{verify pass}($c$) occurs, the FSM transitions to state $q_3$. 
        \item[$\bm{\delta(q_2, f) = q_5}$:] Conversely, if \textbf{verify fail} ($f$) occurs, the FSM transitions to state $q_5$ with the detailed failed reasons. These reasons include explanations for the reproduction failure and suggestions for optimization.
        \item[$\bm{\delta(q_3, d) = q_4}$:] In \textit{(External-Verify)} state $q_3$, an independent external referee determines whether the error message reproduces the issue, called referee verification. If \textbf{verify pass}($d$) occurs, the FSM transitions to the final \textit{(Report)} state $q_4$ and exits. Additional external verification helps to improve the quality of the reproduction scripts.
        \item[$\bm{\delta(q_3, g) = q_5}$:] Conversely, if \textbf{verify fail}($g$) occurs, the FSM transitions to state $q_5$ with the detailed failed reasons provided by the referee.
        \item[$\bm{\delta(q_5, h) = q_5}$:] In \textit{(Modify)} state $q_5$, the Reproducer Agent reflects on the reasons for reproduction failure and edits the current reproduction script. After editing, the patch is first checked, called patch check. Besides the format checking, the patch check ensures that only the reproduction scripts can be modified to keep the process streamlined. If the \textbf{patch check fail}($h$) occurs, the complete reproduction script for reference will be provided and the state remains $q_5$. 
        \item[$\bm{\delta(q_5, i) = q_5}$:] The Reproducer Agent leverages external syntax tools to verify the reproduction script, called syntax check. If the \textbf{syntax check fail}($i$), the state remains $q_5$.
        \item[$\bm{\delta(q_5, j) = q_1}$:] If the script passes all the checks, the FSM transitions back to state $q_1$ with the \textbf{modified reproduction script}($j$).
        \item[$\bm{\delta(q_5, k) = q_6}$:] If the edits number reaches the limit, called \textbf{over edit times}($k$) in state $q_5$, the process transitions to state $q_6$. A straightforward idea is to generate more diverse reproduction scripts from scratch, rather than struggle in refinement.
        \item[$\bm{\delta(q_6, l) = q_0}$:] In \textit{(Restart)} state $q_6$, the Reproducer Agent summarizes the reasons for failure based on the modification history called \textbf{failure experience}($l$). Then the FSM transitions to state $q_0$ and restart. The accumulated failure experiences can serve as references to avoid repeating mistakes.
    \end{description}
\end{itemize}

Overall, we propose the FSM framework to instruct the behavior of the Reproduce Agent through multiple feedback. Specifically, static and dynamic information from ~\textbf{Error Message} and \textbf{Syntax Check} guide \framework in analyzing reproduction situations and take targeted modifications. \textbf{Patch Check} and \textbf{Command Check} direct \framework to focus on optimizing reproduction scripts, thereby avoiding the entanglement of subtasks as shown in Figure~\ref{fig:motivation}(b). \textbf{Referee Verification} introduces external checks to prevent unreasonable verifications as illustrated in Figure~\ref{fig:motivation}(d). Additionally, \textbf{failure experience} and the restart enable \framework to draw from historical experiences, explore various reproduction ways, and avoid repetitive modifications as noted in Figure~\ref{fig:motivation}(d).

\subsubsection{Ranking and Selection}

% \wxc{We extract all reproduction scripts generated during the FSM process and prioritize their submission as follows: (1) \textbf{Double Verification}: Submit the script first if it is confirmed by both the Reproducer Agent and the Referee Verification. (2) \textbf{Single Verification}: If no scripts pass double verification, submit the script next based on confirmation from the Reproducer Agent alone. (3) \textbf{Modification History}: If no scripts meet the above conditions, submit the script that has been most recently modified.}
During the FSM process, we extract reproduction scripts generated at each restart and prioritize them to select the only reproduction script: \textbf{(1) Double-verified Script:} Select the script if it passes both self-verify and external-verify. \textbf{(2) Single-verified Script:} If no scripts pass double verification, select the script that passes self-verify alone. \textbf{(3) Last-modified Script:} If no scripts meet the above criteria, select the most recently modified script.

\section{EXPERIMENTAL SETUP}
\label{sec:evaluation}
In this section, we evaluate the proposed \framework and aim to answer the following research questions (RQs):

\begin{enumerate}[label=\bfseries RQ\arabic*:,leftmargin=.5in]
    
    \item How does \framework perform in the bug reproduction script generation from issue descriptions compared with different methods?
    \item What is the influence of different modules on the performance of \framework?
    \item How do the reproduction scripts generated by \framework improve Agentless's resolved rate?
    \item How do the different hyper-parameters impact the performance of \framework?
    \item What is the distribution of bugs successfully reproduced by \framework regarding issue lengths, time, and project?
    
\end{enumerate}

\subsection{Datasets}
To answer the questions above, we choose the popular SWE-bench Lite~\cite{swebench} dataset to evaluate \framework. SWE-bench Lite consists of 300 real-life GitHub issues from 12 repositories, consisting of issue descriptions and the corresponding open-source repositories. 
% More details of SWE-bench lite can be seen at \textit{https://www.swebench.com/lite.html}. 

\subsection{Comparative Methods}

To evaluate the performance of our framework, we compare two types of bug reproduction script generation methods: LLM-based and Agent-based.

\subsubsection{LLM-based methods}
\begin{itemize}
    \item \textbf{ZEROSHOTLITE} prompts the LLM with the issue description and instructions to generate a new script.
    \item \textbf{ZEROSHOTAGENT} is similar to ZEROSHOTLITE but includes the code context retrieved by the Searcher Agent. 
    \item \textbf{ZEROSHOT}~\cite{DBLP:journals/corr/abs-2406-12952} prompts the LLM with the issue description, related code context retrieved using
BM25~\cite{DBLP:journals/ftir/RobertsonZ09/BM25}, and instructions to generate patch files in unified diff format.
    \item \textbf{ZEROSHOTPLUS}~\cite{DBLP:journals/corr/abs-2406-12952} is similar to ZEROSHOT but leverages adjusted diff format, which allows entire functions or classes to be inserted, replaced, or deleted. 
    \item \textbf{LIBRO}~\cite{DBLP:conf/icse/KangYY23} first generates multiple proposal tests based on the bug report, then it runs all generated tests and selects the test whose error message is most relevant to the issue. 
\end{itemize}

\subsubsection{Agent-based methods}

\begin{itemize}
    \item \textbf{SWE-AGENT}~\cite{DBLP:journals/corr/abs-2405-15793} comprises several principal components, including search, file viewer, file editor, and context management. It is also allowed to employ system commands and utilities.
    \item \textbf{SWE-AGENT+}~\cite{DBLP:journals/corr/abs-2406-12952} is a variant based on the SWE-AGENT, which is instructed to execute the generated reproduction scripts explicitly.
    \item \textbf{AUTOCODEROVER}~\cite{DBLP:journals/corr/abs-2404-05427} consists of two stages: the context retrieval stage and the patch generation stage. In the first stage, AUTOCODEROVER iteratively searches for issue-related code snippets. In the second stage, it generates patches based on the issue description and code context, retrying until the patch is successfully applied.
    \item \textbf{AgentBaseline} is our designed bug reproduction-based agent based on the existing code agent framework as mentioned in Section~\ref{sec:motivation}.
\end{itemize}
% We replace the system prompt like ``solve the issue'' with ``create bug reproduction scripts according to the issue'' to adapt program repair-based code agents for bug reproduction.

\subsection{Implementation Details}

\textbf{Searcher Agent:} 
The Searcher Agent possesses fundamental capabilities for code retrieval and file viewing. It incorporates essential interfaces, such as `\textit{search\_class}' and `\textit{review\_file}', as well as basic system commands, including `\textit{ls}' and `\textit{grep}'. The Searcher Agent iteratively retrieves buggy code and unit tests until it reports "I have retrieved all relevant contexts" or the maximum number of iterations is reached.

\textbf{Hyber-Parameters of Inference:} The Searcher Agent and the Reproducer Agent are provided access to GPT4o-2024-0513~\cite{gpt4o}. 
% By default, we query the LLM with greedy decoding. 
The sampling temperature for the Searcher Agent is set to 0, with a maximum of 40 retrieval iterations. The sampling temperature for the Reproducer Agent is set to 0.7, with a maximum of 5 restarts and up to 5 edits per restart.

\textbf{Backbone Large Language Models:}
\begin{itemize}
    \item \textbf{GPT3.5-turbo}~\cite{gpt35} is a commercial LLM from OpenAI.
    \item \textbf{GPT4}~\cite{gpt4} is a follow-up version of GPT35.
    \item \textbf{GPT4o}~\cite{gpt4o} is a variant of GPT4 optimized for specific applications. 
\end{itemize}

\textbf{Other Details:} 
We construct environment-completed docker images for each instance based on the open-source project SWE-bench-docker~\cite{swebench-docker}. 
% Each docker image must satisfy two conditions: The P2P(Pass-to-Pass) tests in the repository for each instance must run successfully; after applying the gold patch to the repository, both the P2P and F2P(Fail-to-Pass) tests must be run successfully. 
For each instance, \framework launches a new docker container to automatically generate bug reproduction scripts. To speed up script generation, we turn on 4 processes for parallelized inference.

\subsection{Evaluation Metrics}
We identify four different execution states when running the reproduction scripts in the corresponding repository both before and after resolving issues: Fail-to-Pass ($F2P$), Fail-to-Fail ($F2F$), Pass-to-Pass ($P2P$), Pass-to-Fail ($P2F$). We propose three main metrics to assess the performance of any method in generating bug reproduction scripts:

\textbf{The Fail-to-Pass:} The Fail-to-Pass rate ($F \rightarrow P = \frac{F2P}{F2P+F2F+P2P+P2F}$) measures the portion of bug reproduction scripts where the scripts fail in the current repository but pass when the issue has been resolved (\ie the successful reproduction scripts). The higher the Fail-to-Pass rate, the higher the quality of the reproduction script generated by the method.

\textbf{Fail-to-Pass of Fail-to-Any Rate:} This rate (\metricA = $ \frac{F2P}{F2P + F2F}$) measures the proportion of reproduction scripts that successfully execute after the issue is resolved, among those that fail to execute before the issue is resolved.

\textbf{Fail-to-Pass of Any-to-Pass Rate:} This rate (\metricB = $\frac{F2P}{F2P + P2P)}$) measures the proportion of reproduction scripts that fail to execute before the issue is resolved, among those that successfully execute after the issue is resolved.

\section{EXPERIMENTAL RESULTS}
\label{sec:experimental_result}
\subsection{RQ1: Effectiveness of \framework in General Bug Reproduction}

\newcommand{\cross}{\textcolor{red}{\textbf{\XSolidBrush}}}
\newcommand{\tick}{\textcolor{darkgreen}{\Checkmark}}

\begin{table}[t]
    \caption{Comparison results between \framework, the LLM-based methods, and the Agent-based methods on the SWE-bench Lite. The `Retrieve' column indicates if the method retrieves code from the repository, the `Execute' column indicates if the method can execute and then refine the script, and the 'Create' column indicates if the method is instructed to create and edit a new script instead of modifying an existing one. In the $\widehat{F \rightarrow P} $ column, we assume that these methods can successfully reproduce all the filtered instances.}
    \footnotesize
    \centering
    \setlength{\tabcolsep}{0.3mm}
    \begin{tabular}{cl|c|ccc|cc|ccc|c}
    \toprule
    \rowcolor[HTML]{DEDEDE}
       & \textbf{Method} & \textbf{LLM Model} &\textbf{Retrieve} & \textbf{Execute} & \textbf{Create} & $F \rightarrow X$ & $X \rightarrow P$ & $F \rightarrow P$ & \metricA & \metricB & $\widehat{F \rightarrow P}$\\ \midrule
      \textbf{\multirow{9}{*}{LLM-based}} & ZEROSHOT & GPT4 & \tick &\cross  & \cross 
& 6.7
& 9.9
% & 9.5
& 0.4
& 6.0
& 4.0 &16.0 \\ 
       & ZEROSHOTPLUS & GPT4&  \tick & \cross & \cross
& 32.0
& 51.4
% & 45.1
& 6.3
& 19.7
& 12.3 & 21.0\\ 
       & LIBRO & GPT4& \cross & \cross & \cross
& 36.8
& 51.8
% & 42.7
& 9.1
& 24.7
& 17.6 & 23.3\\ 
       & ZEROSHOTLITE & GPT3.5-turbo&\cross  &\cross  &  \tick
& 62.3
& 43.0
% & 35.7
& 7.3
& 11.7
& 17.0 & -\\ 
       & ZEROSHOTLITE & GPT4&\cross  &\cross &  \tick
& 53.3
& 50.0
% & 43.3
& 6.7
& 12.6
& 13.4 & -\\ 
       & ZEROSHOTLITE & GPT4o&\cross  &\cross &  \tick
& 42.3
& 61.0
% & 54.3
& 6.7
& 15.8
& 11.0 & -\\ 
      & ZEROSHOTAGENT & GPT3.5-turbo& \tick &\cross &  \tick
& 39.0
& 39.0
% & 33.0
& 6.0
& 15.4
& 15.4 & -\\ 
       & ZEROSHOTAGENT & GPT4& \tick &\cross &  \tick
& 55.0
& 49.7
% & 42.0
& 7.7
& 14.0
& 15.5 & -\\ 
       & ZEROSHOTAGENT & GPT4o& \tick &\cross&  \tick 
& 54.3
& 53.0
% & 42.7
& 10.3
& 19.0
& \cellcolor{gray!25}19.4 & -\\ 
       \midrule
       \textbf{\multirow{4}{*}{Agent-based}} & SWE-AGENT & GPT4& \tick &\tick  &\cross
& 36.4
& 70.0
% & 60.1
& 9.9
& \cellcolor{gray!25}27.2
& 14.1 & 24.0\\
       & SWE-AGENT+ & GPT4& \tick &\tick &\cross
& 34.0
& 72.0
% & 60.9
& \cellcolor{gray!25}11.1
& \cellcolor{gray!45}32.6
& 15.4 & 25.0\\ 
       & AUTOCODEROVER & GPT4& \tick &\tick &  \cross
& 38.3
& 46.2
% & 38.7
& 7.5
& 19.6
& 16.2 & 22.0\\ 
        & AgentBaseline & GPT4o& \tick &\tick &  \cross
& 48.0
& 41.3
% & 28.3
& \cellcolor{gray!45}13.0
& 27.1
& \cellcolor{gray!45}31.5 & -\\ \midrule
        & \textbf{\framework} & GPT4o& \tick &\tick &  \tick
& 90.0
& 45.0
% & 9.0
& \cellcolor{gray!70}\textbf{36.0}
& \cellcolor{gray!70}\textbf{40.0}
& \cellcolor{gray!70}\textbf{80.0} & -\\ 
    \bottomrule
    \end{tabular}
    \label{tab:rq1_1}
\end{table}

To answer RQ1, we conduct a comprehensive comparative analysis against five LLM-based methods and four Agent-based methods across three metrics. The experimental results are shown in Table~\ref{tab:rq1_1}\footnote{We follow the experimental results of ZEROSHOT, ZEROSHOTPLUS, LIBRO, SWE-AGENT, SWE-AGENT+, and AUTOCODEROVER by Niels et al~\cite{DBLP:conf/icse/KangYY23}. The evaluation of these methods is based on a subset of SWE-Bench lite. Since the filtered instances are not publicly available and the difficulty of issue reproduction is generally averaged, the overall reproduction rate shows minimal fluctuation. Additionally, in the $\widehat{F \rightarrow P} $ column, we assume that these methods can successfully reproduce all the filtered instances.}.

\textbf{The proposed \framework consistently exhibits superior performance compared with the baseline methods.} As shown in Table~\ref{tab:rq1_1}, \framework outperforms all the baseline methods across three metrics.
% and achieves the highest performance.
% in each. 
Specifically, \framework improves by 23.0\%, 7.4\%, and 48.5\% over the best baseline method across the three metrics, respectively. Compared to the average performance of current methods, \framework demonstrates great improvements of 28.2\%, 21.0\%, and 64\% in three metrics. Regarding the $\widehat{F \rightarrow P} $ metric, \framework also surpasses the best baseline by 11.0\%.

\textbf{The feedback information aids in reproducing the bug.} We discover that agent-based methods, such as SWE-AGENT+, AgentBaseline, and \framework, consistently outperform LLM-based methods. Specifically, agent-based methods achieve top-3 results in 8/9 metrics. When considering average performance, agent-based methods show improvements of 8.8\%, 13.9\%, and 17.5\% over LLM-based methods across the three metrics. These results highlight the advantages of agent-based methods, which can leverage various external tools and system commands to self-optimize and adapt based on feedback.

\textbf{Generating a new script is more effective than modifying an existing one}. For LLM-based methods, ZEROSHOTLITE and ZEROSHOTAGENT are tasked with generating a new reproduction script named `\textit{reproduce.py}', while ZEROSHOT and ZEROSHOTPLUS are required to modify or rewrite an existing script. As shown in Table~\ref{tab:rq1_1}, the former approaches achieve an average improvement of 1.9\% over the $F \rightarrow P$ metric compared to the latter. The results reveal that LLMs continue to face challenges in generating patches that meet the required specifications.

\textbf{The impact of code context on LLM-based methods varies across different LLMs.} The only difference between ZEROSHOTLITE and ZEROSHOTAGENT is the inclusion of additional code context. The results indicate that code context improves performance by 1.0\% and 3.6\% on the $F \rightarrow P$ metric for GPT-4 and GPT-4o, respectively, while performance decreased by 1.3\% for GPT3.5-turbo.

\begin{tcolorbox}
 \textbf{Answer to RQ1:} \framework outperforms all the baseline methods in all metrics, achieving 23.0\%, 7.4\%, and 48.5\% improvements over the best baseline, respectively. Besides, we conclude three findings: (1) the feedback information aids in reproducing the bug; (2) generating a new script is more effective than modifying an existing one; (3) the impact of code context on LLM-based methods varies across different LLMs.
\end{tcolorbox}

\subsection{RQ2: Effectiveness of Different Modules in \framework}

\begin{table}[tbp]
    \caption{Impact of the \moduleA(\ie CCC) and the \moduleB (\ie FMO) on the performance of \framework. PASS@1 and PASS@5 settings refer to the evaluation of one or five generated reproduction scripts, respectively. Time represents the average time required to generate a single reproduction script, measured in seconds.}
    \footnotesize
    \centering
    \setlength{\tabcolsep}{2.1mm}
    \begin{tabular}{c|ccc|ccc|c}
    \toprule
    \rowcolor[HTML]{DEDEDE}
      & \multicolumn{3}{c|}{$PASS@1$} & \multicolumn{3}{c|}{$PASS@5$} & \\ 
     \rowcolor[HTML]{DEDEDE}
      \textbf{Module} &  $F \rightarrow P$ & \metricA & \metricB & $F \rightarrow P$ & \metricA & \metricB & Time(s)\\ \midrule
       \textbf{w/o CCC}
& 31.7 \color{darkgreen}{$\downarrow$ 4.3}
& 34.9 \color{darkgreen}{$\downarrow$ 5.1}
& 79.8 \color{darkgreen}{$\downarrow$ 0.2}
& 40.0 \color{darkgreen}{$\downarrow$ 3.7}
& 41.8 \color{darkgreen}{$\downarrow$ 3.7}
& 62.8 \color{darkgreen}{$\downarrow$ 4.0}
& 83
\\
\midrule
        \textbf{w/o FMO}
& 12.3 \color{darkgreen}{$\downarrow$ 23.7}
& 21.8 \color{darkgreen}{$\downarrow$ 18.2}
& 31.4 \color{darkgreen}{$\downarrow$ 48.6}
& 31.0 \color{darkgreen}{$\downarrow$ 12.7}
& 35.4 \color{darkgreen}{$\downarrow$ 10.1}
& 34.6 \color{darkgreen}{$\downarrow$ 32.2}
& 95
\\
\midrule
       \textbf{w/o Modify-in-FMO} 
& 26.0 \color{darkgreen}{$\downarrow$ 10.0}
& 30.7 \color{darkgreen}{$\downarrow$ 9.3}
& 65.0 \color{darkgreen}{$\downarrow$ 15.0}
& 37.7 \color{darkgreen}{$\downarrow$ 6.0}
& 40.2 \color{darkgreen}{$\downarrow$ 5.3}
& 56.8 \color{darkgreen}{$\downarrow$ 10.0}
& 16
\\
\midrule
       \textbf{w/o Restart-in-FMO} 
& 33.3 \color{darkgreen}{$\downarrow$ 2.7}
& 36.2 \color{darkgreen}{$\downarrow$ 3.8}
& 81.3 \color{red}{$\uparrow$ 1.3}
& 48.7 \color{red}{$\uparrow$ 5.0}
& 50.2 \color{red}{$\uparrow$ 4.7}
& 67.0 \color{red}{$\uparrow$ 0.2}
& 198
\\
\midrule
       \textbf{w/o Rank-in-FMO} 
& 16.7 \color{darkgreen}{$\downarrow$ 19.3}
& 49.0 \color{red}{$\uparrow$ 9.0}
& 86.2 \color{red}{$\uparrow$ 6.2}
& 19.0 \color{darkgreen}{$\downarrow$ 24.7}
& 53.8 \color{red}{$\uparrow$ 8.3}
& 82.6 \color{red}{$\uparrow$ 15.8}
& -       
\\
\midrule
       \textbf{w/o Referee-in-FMO}
& 35.0 \color{darkgreen}{$\downarrow$ 1.0}
& 37.8 \color{darkgreen}{$\downarrow$ 2.2}
& 84.0 \color{red}{$\uparrow$ 4.0}
& 41.7 \color{darkgreen}{$\downarrow$ 2.0}
& 43.3 \color{darkgreen}{$\downarrow$ 2.2}
& 61.9 \color{darkgreen}{$\downarrow$ 4.9}
& -
\\
\midrule
       \textbf{\framework} 
& 36.0
& 40.0
& 80.0
& 43.7
& 45.5
& 66.8
& 92
\\
    \bottomrule
    \end{tabular}
    \label{tab:rq2}
\end{table}

To answer RQ2, we explore the effectiveness of different modules on the performance of \framework. Specifically, we study the two involved modules including the \moduleA(\ie CCC) and \moduleB(\ie FMO).

\subsubsection{Comprehensive Context Construction Module.}
To understand the impact of this module, we deploy a variant of \framework without the \moduleA(\ie w/o CCC). It directly employs the context retrieved by the Searcher Agent along with the original issue description.

Table~\ref{tab:rq2} shows the performance of the variant on the $PASS@1$ and $PASS@5$ setting. The addition of the \moduleA yields enhancements of 4.3\%, 5.1\%, and 0.2\% across three metrics in the $PASS@1$ setting. Similarly, in the $PASS@5$ setting, three metrics are improved by 3.7\%, 3.7\%, and 4.0\%, respectively. Overall, the results indicate that the \moduleA enables the Reproducer Agent to focus on a streamlined context window, thereby better analyzing the issue description and referencing issue-related code and its explanations to construct reproduction scripts, ultimately enhancing its ability to reproduce bugs. 

\subsubsection{FSM-based Multi-Feedback Optimization Module.}
To explore the contribution of the \moduleB, we also construct a variant of \framework without the \moduleB(\ie w/o FMO). It creates, executes, and refines the reproduction script freely based on the initial system prompt, without following a standardized process.

As illustrated in Table~\ref{tab:rq2}, the addition of the \moduleB leads to improvements of 23.7\%, 18.2\%, and 48.6\% across three metrics in the $PASS@1$ setting. Likewise, in the $PASS@5$ setting, three metrics show enhancements of 12.7\%, 10.1\%, and 32.2\%, respectively. The results demonstrate that \moduleB greatly boosts performance. This enhancement can be attributed to the module's capability in guiding the Reproducer Agent to make appropriate decisions based on muti-dimension feedback, avoiding unregulated actions, and ensuring a controlled and efficient script generation process.

% \subsubsection{Components in FSM-based Multi-Feedback Optimization Module.}

To further investigate the influence of different components within \moduleB, we design four variants: including those without Modify-in-FMO, Restart-in-FMO, Rank-in-FMO, and Referee-in-FMO. These variants represent generating scripts in a single attempt without execution, modifying scripts continuously until reporting success or exceeding the maximum iterations, only selecting scripts that have passed dual verification, and verifying scripts without external referees, respectively. 

Table\ref{tab:rq2} shows the performances of the four variants. Under the $PASS@1$ setting, the components Modify, Restart, Rank and Referee contribute to an improvement in the $F \rightarrow P$ metric by 10.0\%, 2.7\%, 19.3\%, and 1.0\% respectively. In the $PASS@5$ setting, the performance changes are 6.0\%, -5.0\%, 24.7\%, and 2.0\% respectively. The results reveal that within \moduleB, the Rank component has the most noticeable impact on \framework, indicating that our ranking strategy effectively selects the suitable reproduction script according to predefined rules. The Modify component also exhibits a great influence, suggesting that \framework can iteratively optimize the reproduction scripts based on multiple feedbacks, thereby progressively approaching accurate issue reproduction. The Referee component also boosts \framework performance by providing oversight and recommendations from external referees, which help mitigate the Reproducer Agent's overconfidence. The Restart component improves \framework performance in the $PASS@1$ setting, indicating that accumulating failure experiences aids in exploring diverse reproduction ways. Although the metric improves with the w/o Restart-in-FMO variant in the $PASS@5$ setting, it consumes approximately twice the time as \framework. It is noteworthy that the w/o Rank-in-FMO variant shows notable improvements in the \metricA and \metricB metrics. Selecting only the double-verified reproduction scripts is indeed more precise. However, this variant discards too many other potential reproduction scripts, resulting in a much low $F \rightarrow P$ metric.

\begin{tcolorbox}
 \textbf{Answer to RQ2:} The \moduleA and \moduleB can improve the performance of \framework. The CCC module boosts the $F \rightarrow P$ metric of 4.0\% on average and the FMO module enhances \framework by 18.2\% on average. Moreover, each component within the FMO module is crucial and indispensable.
\end{tcolorbox}

\begin{algorithm}[tpb]
    \footnotesize
    \setstretch{1.0}
    \SetAlgoLined
    \footnotesize
    \SetKwInOut{Input}{Input}
    \SetKwInOut{Output}{Output}
    \SetKwProg{Fn}{Function}{:}{}
    \Input{$All Patches$; // All patches generated by Agentless\\
    $Reproduction Scripts$; // Reproduction scripts generated by \framework}
    \Output{$Selected Patch$; // The selected patch }
    \ForEach{$Patch \in All Patches$}{
            $listF2P \leftarrow \varnothing$;\quad$listDiff \leftarrow \varnothing$;\quad$P2P \leftarrow $ \texttt{P2PCHECK}($Patch$); \\
          \If{$P2P = True$}{
            \ForEach{$Reproduction Script \in Reproduction Scripts$}{
            $F2P \leftarrow $ \texttt{F2PCHECK}($Patch$);\quad $Diff \leftarrow $ \texttt{DIFFCHECK}($Patch$); \\
            \uIf{$F2P = True$}{
                $lisF2P.add(Patch)$
            }
            \ElseIf{$Diff = True$}{
                $lisDiff.add(Patch)$
            }
            }
          }
          \uIf{$listF2P \neq \varnothing$}{
            $Selected Patch = \texttt{FrequencyVote}(listF2P)$
          }
          \uElseIf{$listDiff \neq \varnothing$}{
            $Selected Patch = \texttt{FrequencyVote}(lisDiff)$
          }
          \Else{
            $Selected Patch = \texttt{AgentlessVote}(All Patches)$
          }
    }
    \Return{$Selected Patch$}\\
    \caption{Patch Selection via \framework's Generated Reproduction Scripts}
    \label{algorithm1}
\end{algorithm}

\subsection{RQ3: Agentless's Resolved Rate Improvement via \framework}
To answer RQ3, we explore the impact of reproduction scripts generated by \framework on the resolved rate of Agentless~\cite{DBLP:journals/corr/abs-2407-01489}. Agentless is a straightforward method for automatically resolving software issues, consisting of two stages: localization and repair. Agentless leverages LLM to generate multiple potential patches, applying
 regression testing~\cite{DBLP:conf/issre/WongHLA97} to filter out those failing the existing tests in the repository, normalizing the rest with Abstract Syntax Trees~\cite{ast} (AST), and selecting the most frequent patch through voting. 

We design a simple Algorithm~\ref{algorithm1} to integrate the execution information of reproduction scripts with Agentless's patch selection method. The algorithm first filters out patches that fail the existing regression tests (Line 2-3). For each candidate patch, we execute all reproduction scripts generated by \framework before and after applying the candidate patch. If the execution status changes from failure to pass, the patch is added to the \textit{listF2P} (Line 6-7). If the execution outputs differ, the patch is added to the \textit{listDiff} (Line 8-9). Finally, the algorithm selects the most frequent patch in \textit{listF2P} through voting (Line 13-14), then the most frequent in \textit{listDiff}  (Line 15-16), or applies Agentless's selection method to all patches if none meet the criteria ((Line 17-18). A straightforward consideration is that a patch triggering more execution status changes in reproduction scripts (from failure to success) is likely a potential correct patch. Furthermore, a patch causing more output changes in reproduction scripts indicates an alteration in the test execution path. Additionally, we explore the effectiveness of LLM voting as an alternative approach.

As depicted in Table~\ref{tab:rq3}, our designed algorithm achieves a resolved rate of 30.7\%, with a relative improvement of 12.5\% and 31.8\% compared to the Agentless and LLMVote, respectively. The diverse reproduction scripts generated by \framework contain rich execution information, making patch selection more precise and interpretable.

\begin{figure}[t]
    \centering
    \begin{minipage}[b]{0.46\textwidth}
        \centering
        \begin{table}[H]
            \renewcommand{\arraystretch}{1.0}
            \caption{Agentless’s resolved rate improvement via the reproduction scripts generated by \framework.}
            \centering
            \setlength{\tabcolsep}{3mm}
            \begin{tabular}{l|cc}
            \toprule
            \rowcolor[HTML]{DEDEDE}
              \textbf{Method} & \textbf{$\%Resolved$} & \textbf{$Number$}\\ \midrule
              \textbf{Agentless} & 27.3  & 82 \\ \midrule
              \multirow{2}{*}{\textbf{LLMVote}} & 23.3 & 70 \\ 
              & \color{darkgreen}{$\downarrow$ 4.0} &  \color{darkgreen}{$\downarrow$ 12} \\ \midrule
              \multirow{2}{*}{\textbf{\framework}} & 30.7 & 92 \\
              & \color{red}{$\uparrow$ 3.4} &  \color{red}{$\uparrow$ 10} \\ 
            \bottomrule
            \end{tabular}
            \label{tab:rq3}
        \end{table}
    \end{minipage}
    \hspace{0.5cm}
    \begin{minipage}[b]{0.46\textwidth}
        \centering
        \begin{table}[H]
        \caption{Influence of the Searcher Agent's context retrieval capabilities to \framework.}
        \small
        \centering
        \setlength{\tabcolsep}{0.2mm}
        \begin{tabular}{l|cccc}
        \toprule
        \rowcolor[HTML]{DEDEDE}
           \textbf{Method} & $F \rightarrow P$ & \metricA & \metricB \\ \midrule
             AgentBaseline & 13.0 & 27.1 & 31.5 \\ 
             AgentBaseline-BM25 &  14.3 & 24.8 & 33.6 \\ 
             AgentBaseline-Goldfile& 16.0 & 31.8 & 37.8 \\ \midrule
             \textbf{\framework} 
    & 36.0
    & 40.0
    & 80.0 \\ 
             \textbf{\framework-BM25}  
    & 34.7
    & 37.3
    & 85.3 \\ 
             \textbf{\framework-GoldFile} 
    & 36.3
    & 40.6
    & 77.2 \\
            
        \bottomrule
        \end{tabular}
        \label{tab:discuss}
    \end{table}
    \end{minipage}
\end{figure}

\begin{tcolorbox}
 \textbf{Answer to RQ3:} The reproduction scripts generated by \framework greatly contribute to the enhanced resolved rate of Agentless. Our designed algorithm achieves a resolved rate of 30.7\%, representing a relative improvement of 12.5\% compared to Agentless.
\end{tcolorbox}

\subsection{RQ4: Influence of Hyper-parameters on \framework}

\begin{figure}[t]
	\centering
	\includegraphics[width=\textwidth]{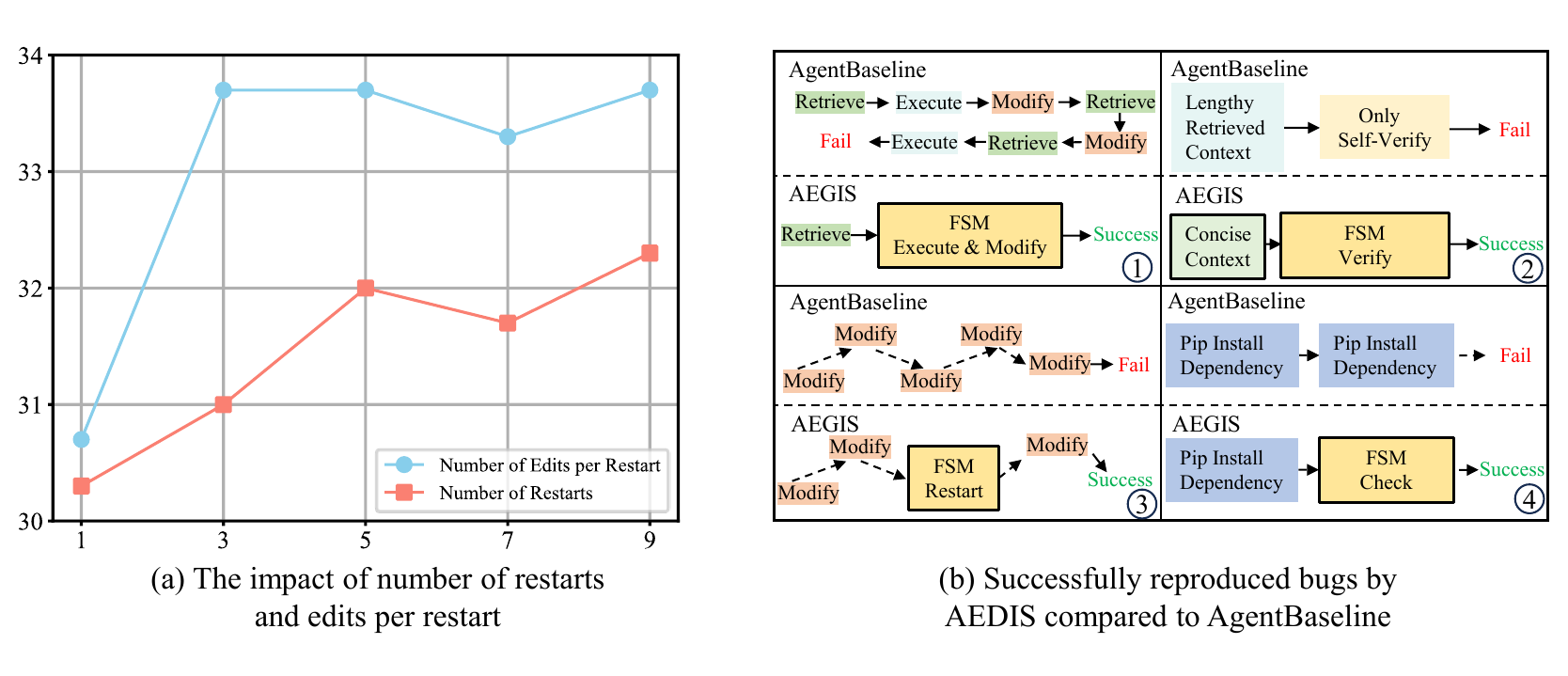}
    \caption{(a) illustrates the impact of the number of restarts and edits per restart on \framework. (b) shows the successfully reproduced bugs by \framework against AgentBaseline.}
\label{fig:hyper}
\end{figure}

To answer RQ4, we explore the impact of different hyper-parameters, including the number of restarts and edits per restart in the \moduleB.

\subsubsection{Number of restarts.}
We experiment on how the number of restarts impacts the performance of \framework. 
Figure~\ref{fig:hyper} illustrates the performance of \framework on the $F \rightarrow P$ metric with different numbers of restarts. As the number of restarts increases(from 1 to 5), the $F \rightarrow P$ metric rises from 30.7\% to 33.7\%. This indicates that as the accumulation of failure experience, \framework can incorporate more suggestions and make divergent attempts. Each restart offers \framework new opportunities to explore different script generation ways while avoiding previous failures. However, as the number of restarts continues to rise(from 7 to 9), the $F \rightarrow P$ metric stabilizes. This implies that after a certain number of restarts, further increases have a limited impact on performance, likely because \framework has already utilized all available failure experiences and cannot derive additional insights.

\subsubsection{Number of edits per restart.}
Similarly, as the number of edits increases, the $F \rightarrow P$ metric ascends from 30.0\% to 32.0\%(from 1 to 5). This demonstrates that \framework can effectively leverage feedback from error messages, syntax checks, patch
checks, and referee verification to iteratively optimize the reproduction script. However, as the number of edits further increases(from 5 to 9), the $F \rightarrow P$ metric tends to stabilize, indicating that further modifications based on the feedback become increasingly challenging. This stabilization may be attributed to \framework reaching an optimization bottleneck, where further improvements require more complex modifications or higher quality feedback. 

\begin{tcolorbox}
 \textbf{Answer to RQ4:} 
 The performance of \framework is influenced by the number of restarts and the number of edits per restart.
 % Our default settings are five restarts and five edits per restart to yield optimal results.
 Our default settings yield optimal results.

\end{tcolorbox}

\begin{figure}[t]
	\centering
	\includegraphics[width=1\textwidth]{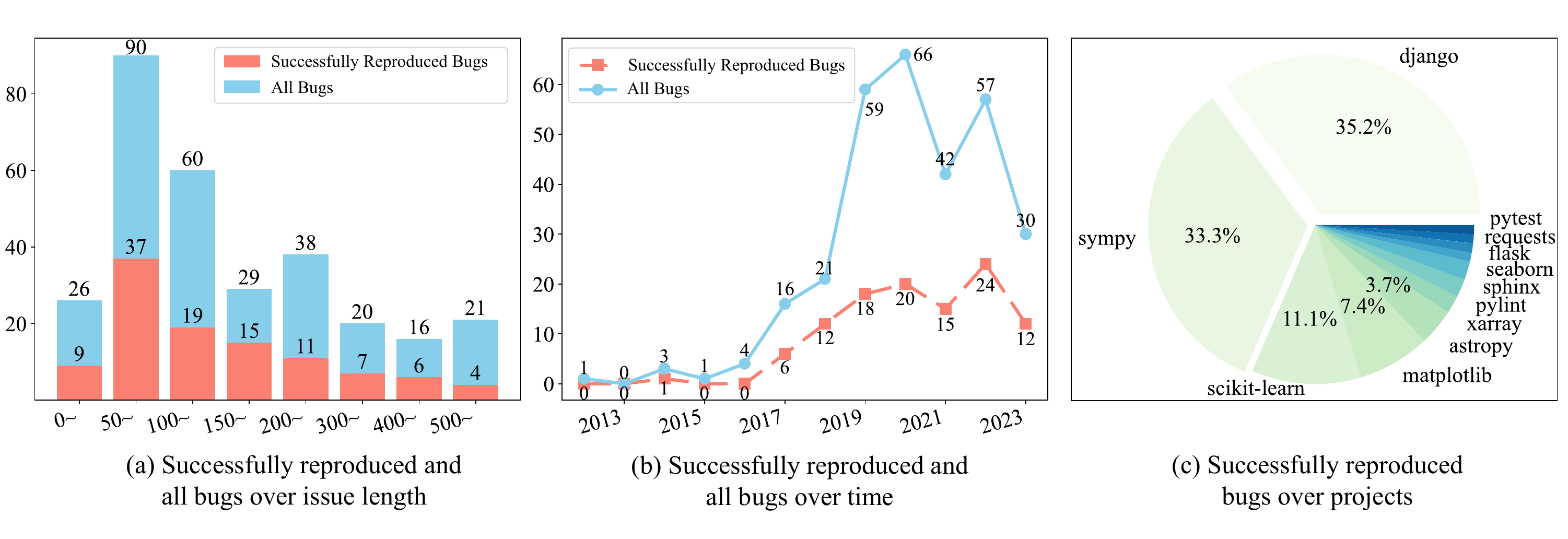}
    \caption{Successfully reproduced bugs over time, issue length, and projects.}
\label{fig:dispaly}
\end{figure}

\subsection{RQ5: Distribution of Successfully Reproduced Bugs by \framework}

To answer RQ5, we present the statistics of successfully reproduced bugs by \framework from different aspects, including issue length, time, and projects.

\textbf{Issue Length:}
Figure~\ref{fig:dispaly}(a) depicts the distribution of successfully reproduced bugs and total bugs in terms of issue length. 
% It shows that instances with issue lengths ranging from 50 to 150 words total 150, accounting for half of the dataset. 
The rates of successfully reproduced bugs are 34.6\%, 39.7\%, and 29.5\% for issues with 0-50 words, 50-200 words, and longer than 200 words, respectively. The result indicates that when the issue descriptions are too brief, the insufficient information hinders the reproduction, while overly long issues make it difficult for agents to understand the issue well.

\textbf{Time:}
Figure~\ref{fig:dispaly}(b) shows the distribution of successfully reproduced bugs and total bugs throughout the past decade. It reveals a consistent upward trend in the number of GitHub issues in SWE-Bench lite, ascending from 1 in 2012 to 57 in 2022. Concurrently, the number of successfully reproduced bugs has also risen, from 1 in 2014 to 24 in 2022. However, the reproduction rate has been fluctuating. While the growth in repository size increases the availability of test cases for reference, it also introduces additional complexity in locating and analyzing bugs.

\textbf{Projects:}
Figure~\ref{fig:dispaly}(c) presents the projects ranked by the number of successfully reproduced bugs. The proportion of successfully reproduced bugs is highest in django~\cite{django} and sympy~\cite{sympy} projects, with 38 and 36 issues, respectively. This trend can be attributed to the more complete testing frameworks and well-structured issue descriptions. In contrast, projects like pytest~\cite{pytest}, requests~\cite{requests}, and flask~\cite{flask} show fewer reproductions. The lower number of requests and flask may be due to their reliance on well-configured network environments. Additionally, many pytest reproduction scripts utilize its proprietary test framework, where successful execution does not necessarily equate to passing tests. 

\begin{tcolorbox}
 \textbf{Answer to RQ5:} The proportion of successfully reproduced bugs first increases and then decreases with the length of the issue description, fluctuating over time. Django and sympy have the highest numbers of successfully reproduced bugs over projects. 
\end{tcolorbox}

\section{DISCUSSION}
\label{sec:discussion}

\subsection{\framework Performance Against Current Challenges}

In Section~\ref{sec:motivation}, we conclude the challenges faced by the code agent in the general bug reproduction task, including entangled subtasks, lengthy retrieved context, and unregulated actions. As shown in Figure~\ref{tab:discuss}(b) We select four cases with different reproduction challenges from those successfully reproduced by \framework to better demonstrate its intelligent capabilities.

The first case(`django-13757') suggests adding unit tests to the existing test files. AgentBaseline only retrieves three rounds of context before attempting to reproduce the bug, frequently returning to context retrieval until the rounds are exhausted. Additionally, AgentBaseline is hindered by modifications across multiple files. In contrast, \framework is restricted to modifying only the bug reproduction script and separates the context retrieval task from the script optimization task. By decoupling these subtasks, \framework can focus more effectively on the current objective, thereby improving performance.

In the second case(`flask-4045'), AgentBaseline retrieves over 400 lines of context and ultimately uses `print' statements to output errors instead of the execution error when constructing the reproduction script. The lengthy context hinders AgentBaseline's understanding, and the final reproduction is only self-verified. \framework's \moduleA helps to distill relevant information from the context for bug reproduction, and the referee verification in \moduleB also alleviates such problem.

In the third case(`django-11964'), AgentBaseline encounters a patch format error while editing a file and continues to attempt corrections until the rounds are exhausted. When \framework encounters a patch format error, it provides the complete reproduction script for reference. Additionally, \framework can perform timely restarts within the FSM process, preventing endless erroneous edits and allowing for more diverse attempts.

In the fourth case(`scikit-learn-25570'), AgentBaseline continuously attempts to install dependencies until time runs out. Within the FSM process, \framework performs command checks on installation commands and rejects them, allowing it to focus more on bug reproduction.

% In Section~\ref{sec:motivation}, we select 39 cases for detailed reanalysis, of which 14 cases include complete reproduction information. Among these 14 cases, \framework successfully reproduces 11. Additionally, two cases are manually verified to be non-reproducible in environment-complete dockers. 

% The only case that fails to reproduce is \textit{`astropy-7746'}, as shown in Figure~\ref{tab:discuss}. In this case, \framework does not simply copy the input example provided in the issue description but attempts to understand the issue's semantics (\ie the test should return an empty list or array). Consequently, it leverages \textit{assert isinstance()} and \textit{assert len()} for validation. However, upon executing the input example after resolving the issue, we discover that the execution message(`[array([], dtype=float64), array([], dtype=float64)]') is not merely an empty array but also contains additional information regarding data dimensions and types. This case demonstrates that \framework can extract and comprehend essential information and generate regulated and reasonable scripts, rather than mechanically replicating the input example. Nevertheless, such an execution message clearly exceeds the agent's understanding capabilities.

\subsection{The Impact of the Searcher Agent on \framework}

In this section, we examine the impact of the Searcher Agent's context retrieval capabilities on \framework. We employ BM25~\cite{DBLP:journals/ftir/RobertsonZ09/BM25} to retrieve files related to the issue description or supply the Searcher Agent with the 'GoldFile', leading to two distinct variants. The 'GoldFile' refers to providing the Searcher Agent with the bug file path. Under this configuration, we posit that the Searcher Agent can ideally pinpoint all relevant code.

The superior performance of AEGIS is not reliant on the Searcher Agent. As illustrated in Table~\ref{tab:discuss}, \framework and \framework-BM25 have respectively achieved substantial improvements of 23.0\% and 20.4\% compared to AgentBaseline and AgentBaseline-BM25. However, the performance enhancement of \framework over \framework-BM25 is merely 1.3\%. Furthermore, \framework-GoldFile has shown a modest 0.3\% improvement compared to \framework. This suggests that \framework can iteratively optimize based on multiple feedback, generate anticipated reproduction scripts, and thus exhibit a weak dependency on the completeness of the gathered context.

\subsection{Threats and Limitations}
\textbf{Dataset Validity Concerns}: 
% SWE-Bench lite covers a wide range of real-world challenges, offering a valuable resource for the assessment of software engineering tools and techniques. 
SWE-Bench Lite's scope is limited because it is derived from only 12 popular repositories, and may not cover all issue types and testing scenarios.
Moreover, SWE-Bench lite exclusively includes issues based on Python, thereby excluding other widely used programming languages. However, \framework can be applied to other programming languages. In future research, we intend to extend our investigations to these alternative programming languages.

\textbf{Result Variability}: Due to \framework relying on code agents for context retrieval and issue reproduction, the experimental results exhibit variability. We therefore conduct multiple trials and average the results to obtain a more stable measure.

\section{RELATED WORK}
\label{sec:related}
\subsection{Automatic Bug Reproduction}

LIBRO~\cite{DBLP:conf/icse/KangYY23} is the first work to automatically generate reproductions for general defects. 
They utilize LLMs to generate bug-reproducing tests and employ post-processing to select promising test cases. 
However, LIBRO cannot dynamically execute and modify reproduction scripts, limiting its performance.
Agents designed for program repair have been employed for bug reproduction tasks, often outperforming traditional approaches such as LIBRO. However, Niels et al.~\cite{DBLP:journals/corr/abs-2406-12952} has primarily relied on adapting existing program repair agents, rather than developing agents specifically tailored for bug reproduction. In this paper, we introduce \framework, a novel approach explicitly designed for bug reproduction, which demonstrates superior effectiveness compared to the method proposed in~\cite{DBLP:journals/corr/abs-2406-12952}. While prior research has explored reproducing specific types of bugs using LLMs in targeted domains, such as Android applications~\cite{DBLP:journals/corr/abs-2407-05165, DBLP:conf/icse/FengC24, DBLP:conf/icse/HuangWLW0CHW24} and configuration-triggered bugs~\cite{DBLP:conf/icse/Fu00DZJ0JL24}, our work focuses on reproducing general software defects from reported issues.

Automated bug reproduction is also related to automatic test case generation~\cite{DBLP:journals/pacmse/Yuan0DW00L24, DBLP:conf/sigsoft/ChenHZHDY24, DBLP:journals/tse/SchaferNET24}. The former generates test cases from natural language descriptions, whereas the latter~\cite{auto1, auto2, auto3, DBLP:journals/pacmse/Yuan0DW00L24} typically generates test cases from the code under test. These efforts are generally aimed at improving coverage.

\subsection{Code Agents}
Code agents can accomplish complex tasks by formulating plans, reflecting, and utilizing tools~\cite{DBLP:journals/corr/abs-2308-10848,DBLP:journals/corr/abs-2307-07924, DBLP:conf/acl/Qiao0FLZJLC24}. 
Code agents are designed specifically for software engineering tasks, 
including requirements engineering~\cite{DBLP:journals/corr/abs-2405-03256, DBLP:journals/corr/abs-2310-13976}, code generation~\cite{DBLP:journals/corr/abs-2312-13010, DBLP:journals/corr/abs-2404-02183}, static bug detection~\cite{DBLP:journals/corr/abs-2403-14274, DBLP:journals/corr/abs-2310-08837}, code review~\cite{DBLP:journals/corr/abs-2402-02172, DBLP:journals/corr/abs-2404-18496}, unit testing~\cite{DBLP:conf/sigsoft/ChenHZHDY24, DBLP:journals/corr/abs-2305-04207}, system testing~\cite{DBLP:journals/corr/abs-2311-08649, DBLP:journals/corr/abs-2308-06782}, fault localization~\cite{DBLP:journals/corr/abs-2403-16362, DBLP:journals/corr/abs-2310-16340}, program repair~\cite{DBLP:journals/corr/abs-2304-00385, DBLP:journals/corr/abs-2403-17134}, end-to-end software development~\cite{DBLP:journals/corr/abs-2405-15793, DBLP:journals/corr/abs-2404-05427} and end-to-end software maintenance~\cite{DBLP:journals/corr/abs-2406-01304, DBLP:journals/corr/abs-2406-01422}. 
% There are also studies showing that program repair tasks can be accomplished without using agents, sometimes even outperforming code agents, 
Several studies have demonstrated that program repair tasks can be effectively accomplished without the use of code agents, with some approaches even outperforming agent-based methods,
as demonstrated by Agentless~\cite{DBLP:journals/corr/abs-2407-01489}. 
Among these works, some code agents, such as CodeR~\cite{DBLP:journals/corr/abs-2406-01304}, MASAI~\cite{DBLP:journals/corr/abs-2406-11638}, and SWE-agent~\cite{DBLP:journals/corr/abs-2405-15793}, consider bug reproduction as a part of the overall process. 
% However, these works do not specifically investigate the effectiveness of bug reproduction and do not deeply analyze why certain cases succeed while others fail. Furthermore, they lack a more optimized agent dedicated to this task. Our work is the first to specifically optimize a code agent for the task of bug reproduction.
% One can refer to the survey~\cite{liu2024largelanguagemodelbasedagents} for more information.  
However, these works do not specifically focus on evaluating the effectiveness of bug reproduction or conducting in-depth analyses of the challenges in this task. Additionally, they lack a dedicated and optimized agent tailored to this specific task. Our research is the first to optimize the code agent specifically for bug reproduction, addressing this critical gap in the field.

\section{CONCLUSION}
\label{sec:conclusion}
This paper focuses on the general bug reproduction from issue descriptions and proposes a novel agent-based framework, named \framework. 
\framework consists of a \moduleA for constructing information from unstructured issue descriptions and extracting issue-related code snippets from extensive contexts and a \moduleB to instruct the code agent to leverage multi-dimensional feedback through iterative optimization.
Compared with the state-of-the-art methods, the experimental results validate the effectiveness of \framework. Furthermore, \framework mitigates the existing challenges associated with bug reproduction tasks. In the future, we intend to further evaluate \framework on a broader range of datasets and programming languages for general bug reproduction tasks.

\section*{Data availability}
Our generated reproduction scripts are available at: \textit{{https://anonymous.4open.science/r/AEGIS}}

\bibliographystyle{ACM-Reference-Format}
\bibliography{Citation} 
\end{document}